\documentclass[conference]{IEEEtran}
\ifCLASSINFOpdf
  \usepackage[pdftex]{graphicx}
  \graphicspath{{../pdf/}{../jpeg/}}
  \DeclareGraphicsExtensions{.pdf,.jpeg,.png}
\else
  \usepackage[dvips]{graphicx}
  \graphicspath{{../eps/}}
  \DeclareGraphicsExtensions{.eps}
\fi
\usepackage{amsmath}
\usepackage{algorithmic}
\usepackage{subfig}
\begin{document}
%
\title{Data-driven Analytics for Business Architectures: Proposed Use of Graph Theory}

\author{\IEEEauthorblockN{Lei Huang, Guangjie Ren, Shun Jiang, Raphael Arar, Eric Young Liu}
\IEEEauthorblockA{IBM Research - Almaden\\
San Jose, California 95120\\
Email: \{lei.huang1, eric.young.liu\}@ibm.com, \{gren, sjiang, rarar\}@us.ibm.com}
}


%


\maketitle

\begin{abstract}
Business Architecture (BA) plays a significant role in helping organizations understand enterprise structures and processes, and align them with strategic objectives. However, traditional BAs are represented in fixed structure with static model elements and fail to dynamically capture business insights based on internal and external data. To solve this problem, this paper introduces the graph theory into BAs with aim of building extensible data-driven analytics and automatically generating business insights. We use IBM's Component Business Model (CBM) as an example to illustrate various ways in which graph theory can be leveraged for data-driven analytics, including what and how business insights can be obtained. Future directions for applying graph theory to business architecture analytics are discussed.
\end{abstract} 
\qquad

\qquad

\qquad

\qquad
%
\IEEEpeerreviewmaketitle
\section{Introduction}
In the age of globalization with fierce competition, dynamic marketplaces and changing customer demands, it is critical that business organizations understand their structures and processses, align business strategy and organization's capabilities and investment, detect and reorganize redundant business capabilities (especially after mergers and acquisitions), and recognize business innovations. Consequently, great efforts have been made to design and improve Business Architecture (BA) models for solving those challenges. 

Business Architecture (BA) is a blueprint of the enterprise that provides a common understanding of the organization and is used to align strategic objectives and tactical demands, which articulates the structure of an enterprise in terms of its capabilities, governance structure, business processes, and business information \cite{BADefinition}. In the past few decades, various BA models have been developed, of which some major models including: ArchiMate \cite{Archimate} that is maintained by the Archimate Foundation and approved as technical standard by the Open Group, and can be used to formally describe business operations; Business Architecture Working Group (BAWG) \cite{BAWG} that is founded as a part of the Objected Management Group (OMG) for establishing industry standards, supporting the creation, and alignment of business blueprints; Business Motivation Model \cite{BMM} that is used for establishing, communicating, and managing business plans, and is also a standard of the OMG. Business Process Modeling Notation \cite{BPMN} that is released by OMG for linking business process model design and process implementation; Business Concepts \cite{BC5387077} that is introduced by McDavid in 1996 to model business concerns relevant for information development; Component Business Model (CBM) \cite{IBMCBM} that is developed by IBM and actively applied in IBM Global Business Services (GBS); Enterprise Business Architecture (EBA) \cite{EBA} that is developed by Gartner to optimize business components along with information and technology; Event-Driven Process Chain \cite{ARISEDPC} that is widely used for documentation and enterprise operation analysis; Enterprise Business Motivation Model \cite{BMMBRG} that is developed by Microsoft's enterprise architect Nick Malik with the aim of aligning enterprise actions with objects; and The Open Group Architecture Framework (TOGAF) \cite{TOGAF} is developed and maintained by The Open Group to optimize strategy, structure and operations. 

Generally, those BAs comprises three core dimensions, conceptual model, which is also referred as modeling language that offers modeling constructs that cover the business domains of an enterprise fully or partially; methodology, which describes the processes, techniques for developing Business Architecture models;  and tooling, which provides user-friendly tools with functionalities, such as developing environment, visualization, analysis and so on \cite{Glissman2009ACR}. However, all the three dimensions are mainly for designing and developing the BA models. Once the BA model is built, the structures are fixed with static model elements while all business insights are manually analyzed and captured, even though some BA tools may make the manual work slightly easier. Furthermore, conventional BAs work as the bridge between enterprise strategies and business functionalities, but are insensitive to outside changes (e.g. dynamic marketplaces, variable customer needs, etc.). How to use BAs as an effective connection between internal strategy and outside changes remains a challenge. 

In this paper, we propose the use of graph theory for data-driven analytics with the aim of making BAs more extensible and intelligent so as to provide more business insights. We use CBM as an example to illustrate the methods and applications. Firstly, we introduce definitions, properties and applications of graph theory, and we provide more background about CBM method. Then, we present how the graph theory based data-driven analytics is implemented in CBM and explore what and how business insights can be obtained through the data-driven analytics. We conclude the paper by summarizing our current work, and discussing future directions.

\section{Methods}
\subsection{Graph Theory}
Graph theory is a mathematical structure that can be effectively used to model pairwise relations between objects. Mathematically, a graph can be presented as $G = (V, E)$, where $V = (v_1, v_2, ..., v_n)$ is the node set and $E = (v_i, v_j)$ is the edge set that contains the pairwise relations between nodes. The graph $G$ can be undirected or directed, which depends on whether the relation between node pairs is symmetric. Moreover, weighted graph can be built to indicate the strength of the relations by using different values to weight edges, in which the edge set is $E = (v_i, v_j, w_{i,j})$.

The graph has various forms in different research areas, such social network \cite{NIPS2012_4532}, biological network \cite{Huang2016,leiplosone2016, Huang2016JBSB, Huang2018}, transportation network \cite{7223282} and so on. The study of graph theory covers a broad research topics and methods, which generally can be classified into the following categories: a) topological analysis, such as centralities (e.g. degree, betweenness, closeness etc.), connectivity,  and so on; b) network component analysis, such as network clustering, clique detection and so on, and c) graph evolution and completion including link prediction, network matching, etc. Overall, graph theory has demonstrated capability to solve many practical problems in different areas.

Although some BAs, e.g. as ArchiMate \cite{Archimate}, Business Process Modeling Notation \cite{BPMN}, Event-Driven Process Chain \cite{ARISEDPC}, etc., has adopted graph structure to build the model, they just used the graph as a container and did not further utilize graph theory to get more business insights. Thus, it is of great interest to investigate and study the graph theory's potential capability for business insight analytics. 

\subsection{Integration of Graph Theory and Component Business Model (CBM) }
\subsubsection{Introduction of CBM}
We choose the CBM models as the study case for the graph theory based analytics. The Component Business Model (CBM) is a BA developed and used by IBM Global Business Services to help clients analyze business from multiple perspectives, and clarify a company's focus on strategic, differentiating capabilities, enabling more straightforward prioritization of improvement plans \cite{CBMIntro2004}.
\begin{figure*}[!t]
\centering
\includegraphics[width=5in]{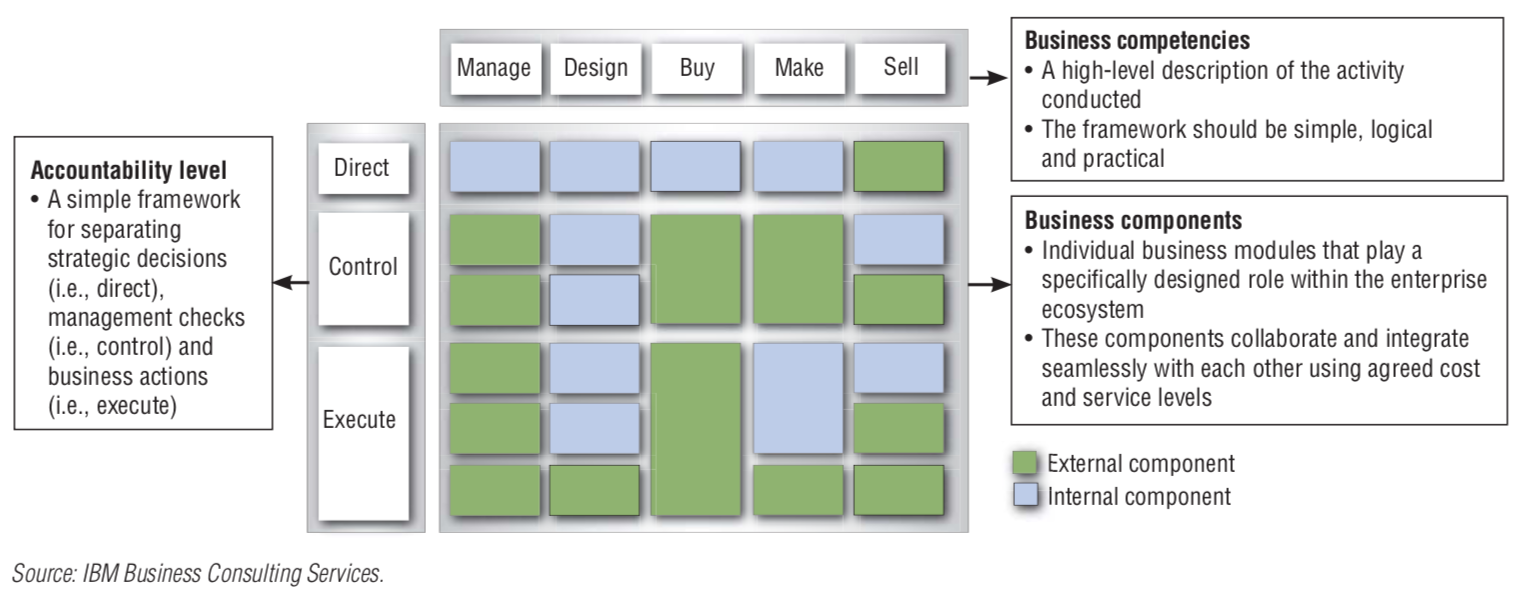}
\caption{Component Business Model Layout.}
\label{fig_cbm}
\end{figure*}
The business component is the fundamental building block of the CBM model, which consists of data, process, people and systems. As shown in Fig.\ref{fig_cbm}, typically, a business component can be further decomposed into several layers. Meanwhile, components are assigned to business competencies, which represent a large business area with skills and capabilities, as well as accountabilities levels, which is used to separate strategic decisions, control mechanisms, and business actions \cite{Glissman2009ACR}. Moreover, the CBM has three business views, (strategic view, financial view, and transformational view) for clear competitive differentiation, investment/cost evaluation, and improvement opportunities.

Although traditional Component Business Model (CBM) provides a concise and straightforward one-page framework for viewing and modeling businesses, the grid structure makes CBM components are internally and externally isolated: internally, we cannot visualize and analyze the relations and influences among business components; externally, components are isolated from outside influences (e.g dynamic marketplaces, changing customer demands, etc.). The intrinsically fixed and limited structure implies the current CBM can only provide manual and static analytics. Although, Jorge Sanz et al. \cite{Sanz2018} tried to build a smart network to connect components, but it did not further explore any data-driven analytics. Therefore, graph theory based data-driven analytics is still a challenge needs to be solved.
\subsubsection{Implementation of Graph Theory based Data-driven Analytics}
\begin{figure}[!t]
\centering
\includegraphics[width=3in]{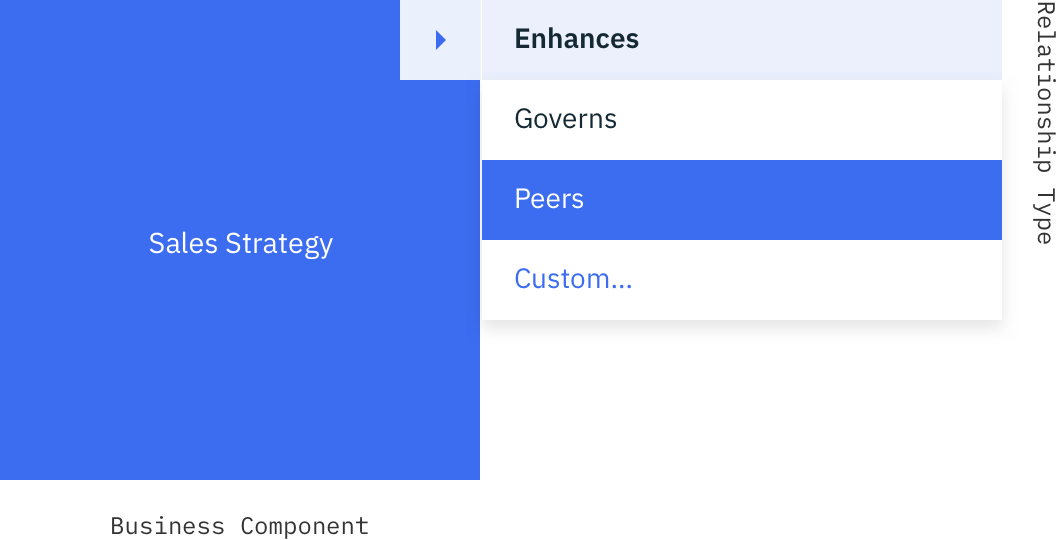}
\caption{Edge types of CBM graph.}
\label{fig_edge_type}
\end{figure}

Similar to the conventional CBM model, we also use the component as the building block for the CBM graph. In additional to the existing attributes (e.g. title, description, processes, etc.), we also add relation types to connect components. As a result, in the graph, nodes are components, edges are relations among components. As shown in the Fig.\ref{fig_edge_type}, when users create or edit a component, they can select a relations type that is used to connect with the other component. We predefined several relation types as default choices for users, while they can also customize their own relation types and assign weights for those relations as well. Thus, the CBM graph contains indirect or direct edges, which depends on different relations types (e.g. governs type is direct, peers type is indirect, etc.); and the Fig.\ref{fig_cbm_visualization}(b) shows an illustrative example of CBM graph. Once the graph is built, then we can do various analytics based on the component's internal attributes, graph structure, and external data.

\section{Data-driven Business Insights}
\subsection{Structure Visualization }
\begin{figure*}
\centering
\subfloat[CBM Model Visualization.]{{\includegraphics[width=3.4in]{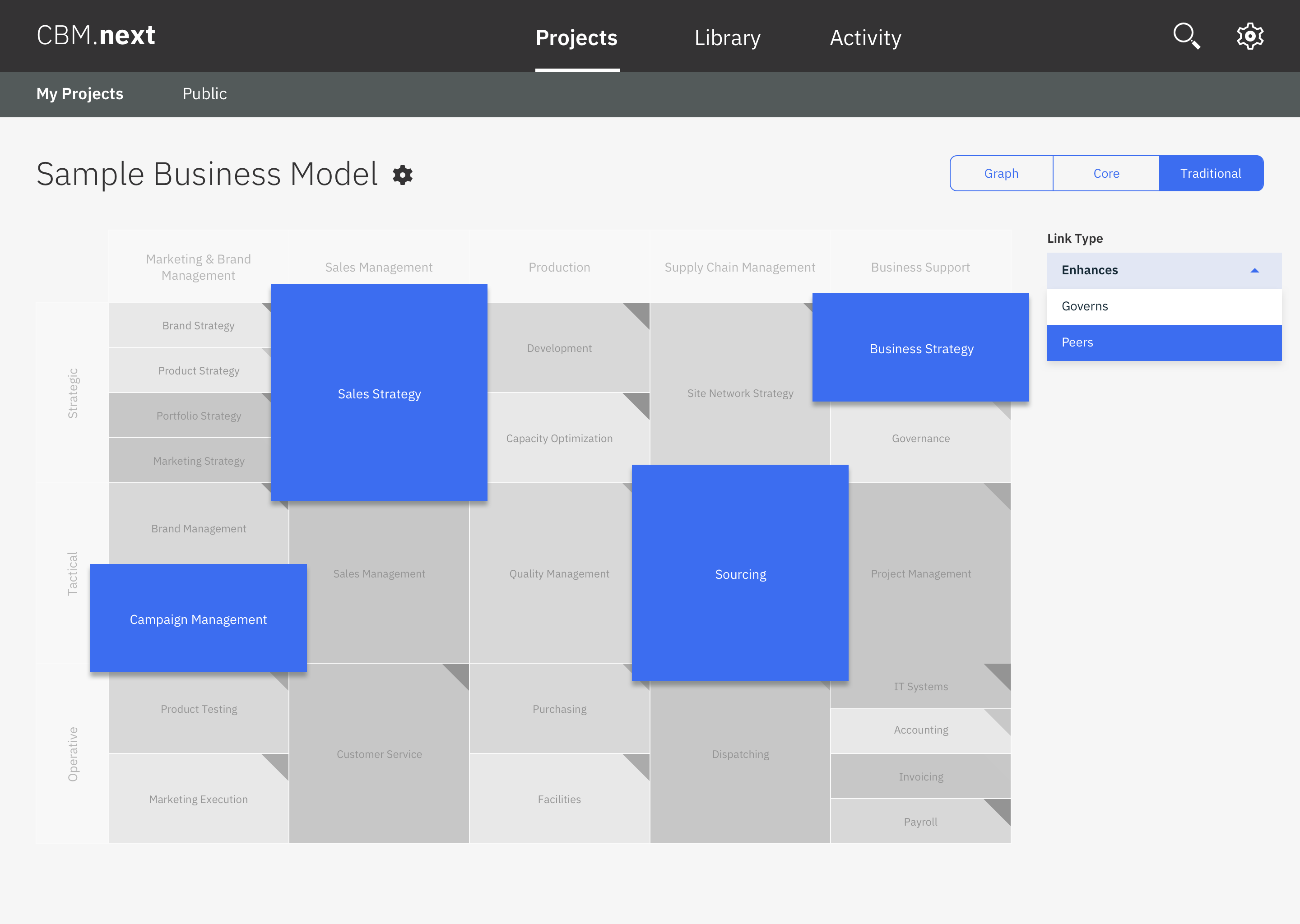} }}%
\quad
\subfloat[CBM Graph Visualization.]{{\includegraphics[width=3.4in]{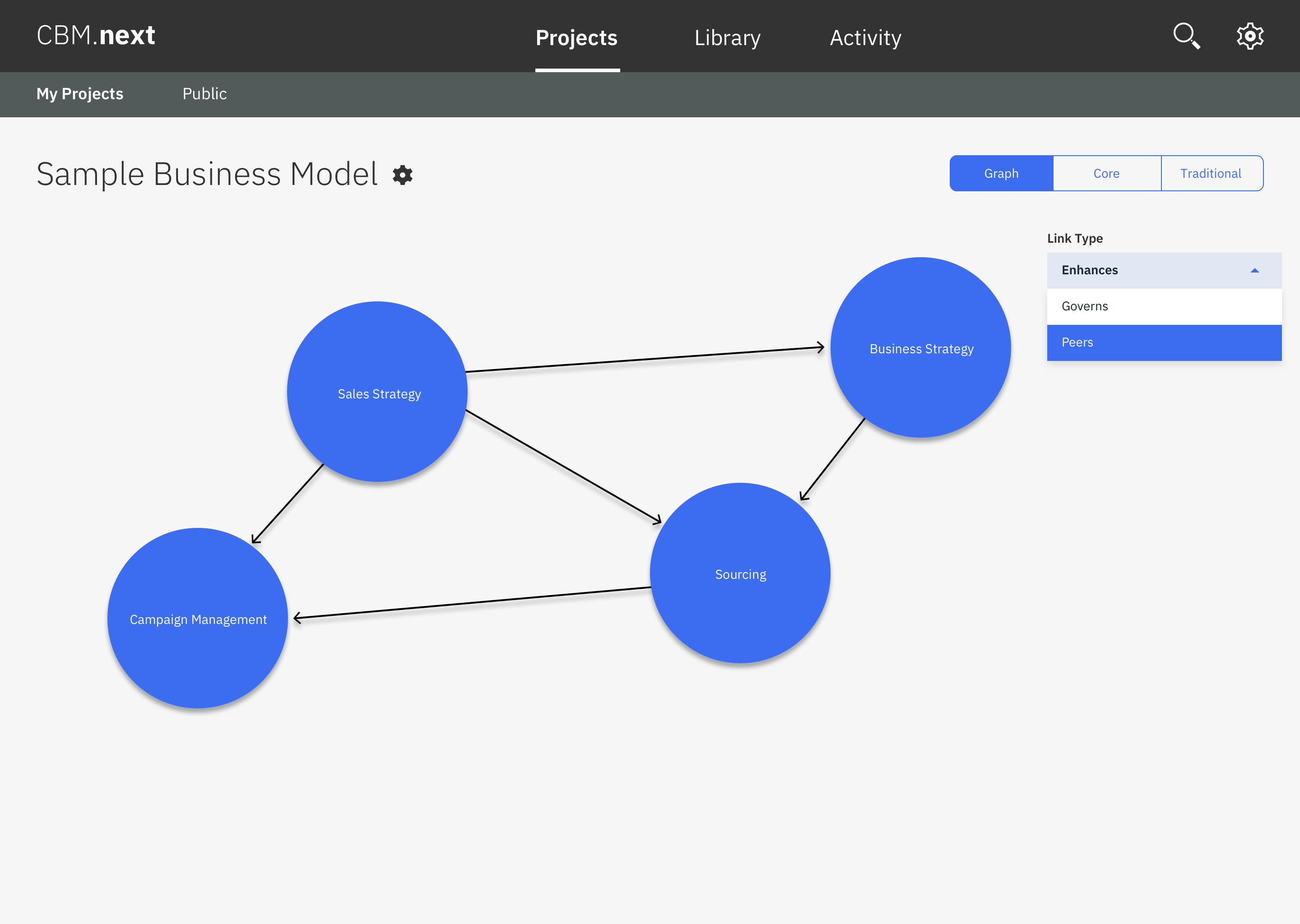} }}%
\caption{CBM Visualization.}%
\label{fig_cbm_visualization}%
\end{figure*}

In the conventional CBM model, all the components are laid in the grid form. Although components can be grouped vertically into competency for high-level description of skills and capabilities, and horizontally into accountabilities for separating strategic decisions, users cannot view the hidden relations among those components, which is actually essential for understanding and improving the CBM model structure and functions.
With the help of CBM graph, users cannot only visualize the overall graph structure but also specific relations among components by selecting edge types, which as shown in the Fig.\ref{fig_cbm_visualization}(a)(b). Meanwhile, by analyzing the components distribution of a specific view, users can also get insights about the relations at competency and accountability level.

\subsection{Business Architecture Analysis based Graph Topological Analysis}
BAs' structure analysis is essential for understanding enterprise organization, prioritizing investment, detecting risk, etc. Although various analytics methods have been adopted by BAs (e.g. conventional CBM uses heap map to indicate importance of component), most of them are based on manual work that could be subjective sometimes. In this section, we would like to discuss and demonstrate how the graph theory can help us intuitively do the data-driven analytics, and get business insights. We present the graph based analytics in a topologically bottom-up sequence.

\begin{figure}[!t]
\centering
\includegraphics[width=3.4in]{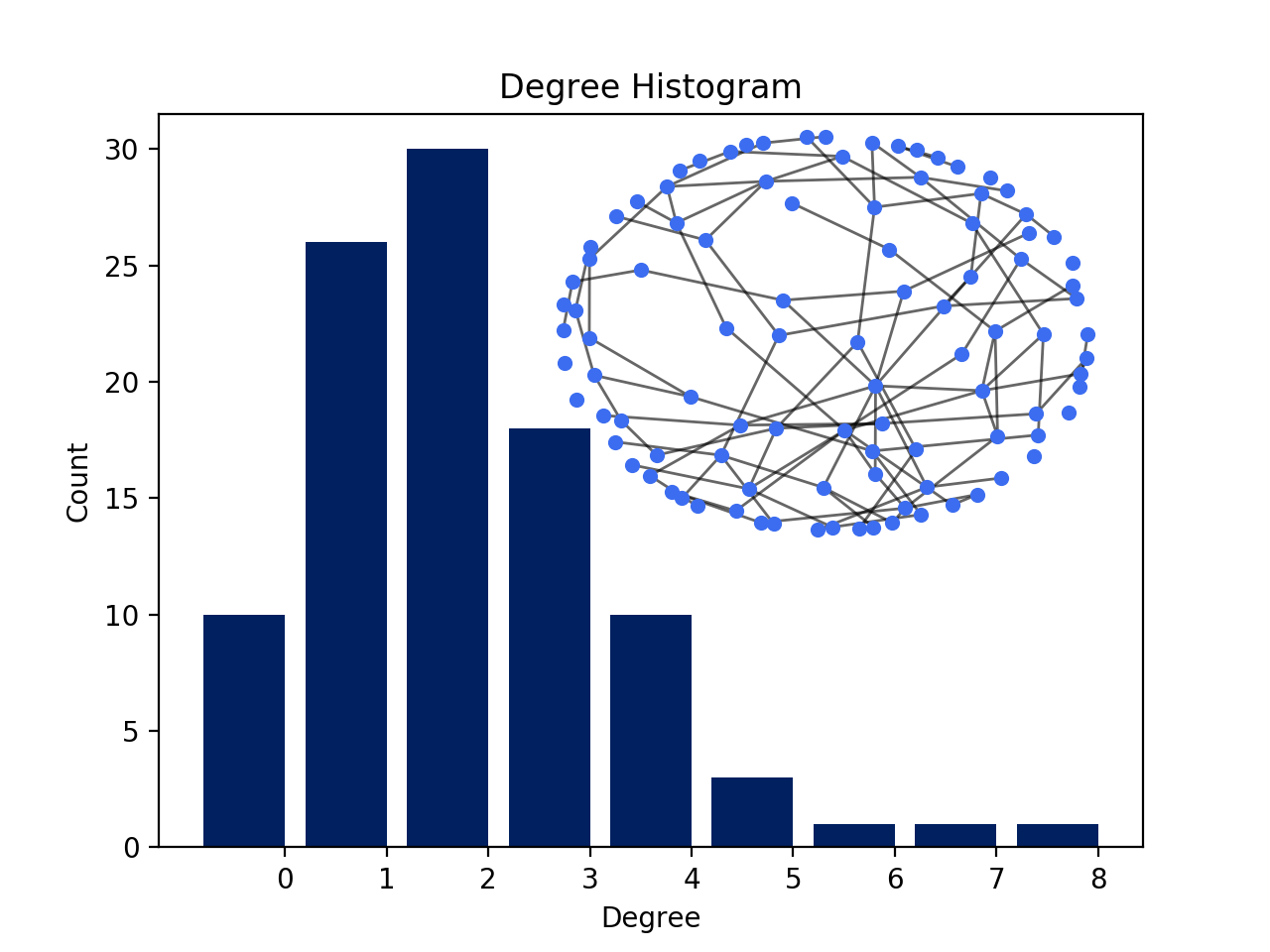}
\caption{The Degree Histogram of Components.}
\label{fig_degree_hist}
\end{figure}

\begin{figure}[!t]
\centering
\includegraphics[width=3.4in]{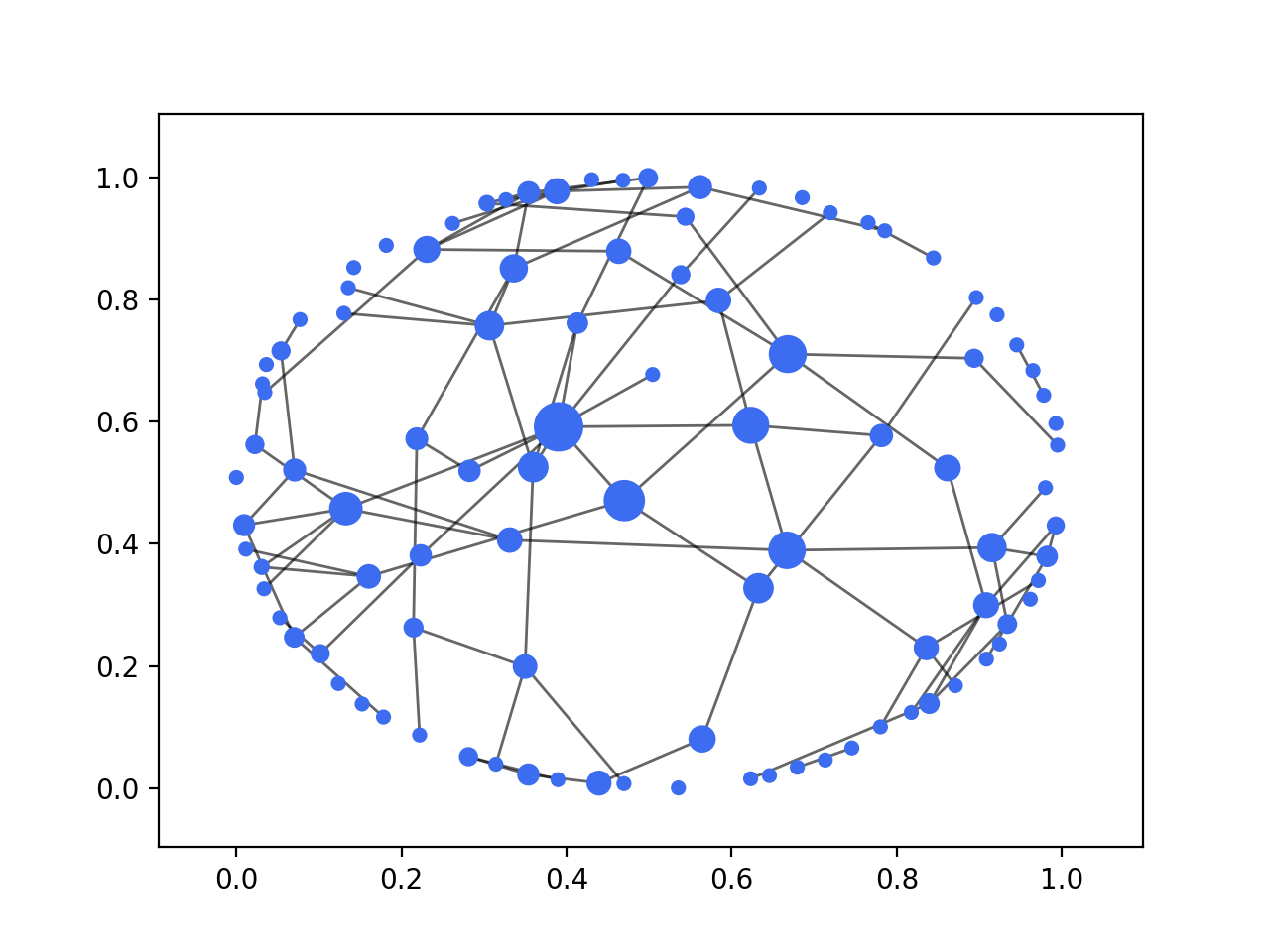}
\caption{The Betweenness Analysis.}
\label{fig_betweenness}
\end{figure}

Firstly, we analyze the importance of business component based on graph centrality. Generally, if a business component plays an important role in a given enterprise, it should more likely be connected with other components. So we can use degree centrality $C_D(v) = deg(v)$ to indicate the component importance, which is defined as the number of links incident upon a node in the graph. We can draw the histogram Fig.\ref{fig_degree_hist} to show whether the business structure is balanced or not. For example, a left concentrated distribution may indicate the enterprise has a loose internal structure, which means the enterprise may need to strengthen the internal communication and collaboration to improve business. On the other hand, a highly right concentrated histogram may indicate there are too many dependencies among business components, which can be the cause of low working efficiency. Therefore, a balanced structure may be the best for business success. Notably, some "trivial" components (e.g. payroll) also have high degree, but they cannot be considered as the major impact for strategic decision. To avoid this problem, we can analyze sub-graph degree centrality by selecting different edge(relation) types. 
Moreover, we can represent the business flow of completing a business goal by the path in the graph. Then we can use the closeness centrality $C(v_i) = \frac{1}{\sum\limits_{j} d(v_i, v_j)}$ to measure the working steps and efficiency, where $d(v_i, v_j)$ is the length of shortest path between components $v_i$ and $v_j$. Meanwhile, for a given component, we can further evaluate the its importance in various business flow by betweenness centrality as shown by the toy example in Fig.\ref{fig_betweenness}, which measures the number of times a node acts as a bridge along the shortest path between two other nodes. Besides, we can also use the edge betweenness centrality to quantify the importance of a business step (between two components) in a business flow. There are many other centralities, such as, eigenvector centrality, PageRank centrality and so on, can help us analyze business structure and get business insights.

\begin{figure}[!t]
\centering
\includegraphics[width=3in]{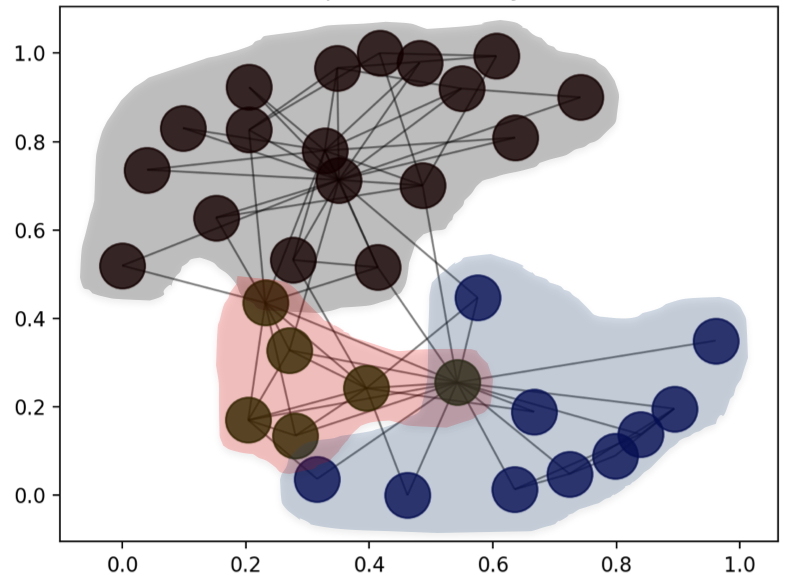}
\caption{CBM Graph Clustering.}
\label{fig_clustering}
\end{figure}

The conventional CBM use competency and accountability to group components, which give us two-dimension view of the business structure. It is of great interest to analyze components from new dimensions. Especially for the large organization, it is important to know what components cross competency and accountability to function closely with each other and form function communities, and how many communities are developed in the organization. By utilizing the network clustering techniques \cite{network_clustering}, such as density based clustering, pattern based clustering and so on, we can easily detect various communities, which is as shown in the Fig.\ref{fig_clustering}

\begin{figure}[!t]
\centering
\includegraphics[width=3in]{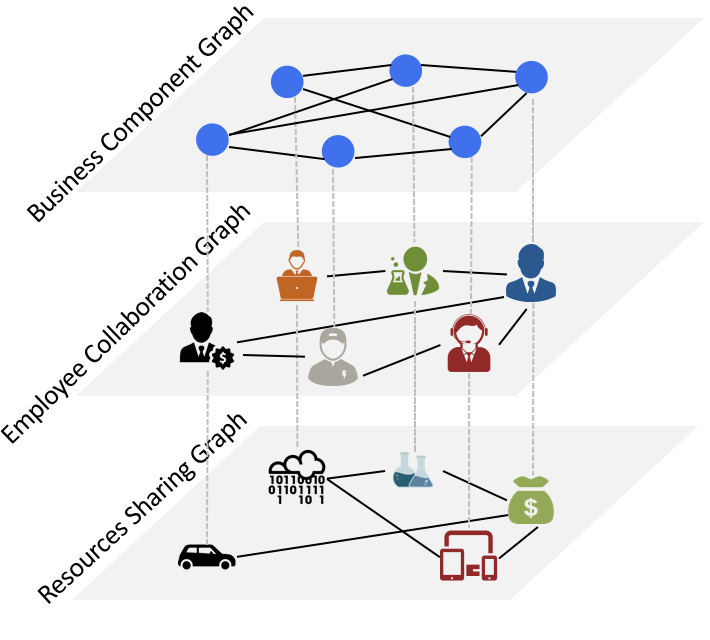}
\caption{Multi-dimension CBM Graph.}
\label{fig_multi_dimen}
\end{figure}

In the above, we discussed the way of using topological and relational features to obtain business insights. To obtain more in-depth insights, it is important to combine those features with underlying features in the components such as data, people, resources, and system. Similarly, we can use edges among business components to connect those underlying features thereby build multidimensional graph with component graph on the top and others laid below, which is as shown in Fig.\ref{fig_multi_dimen}. For example, by connecting people we can get an internal social collaboration network, which help us understand the human resources system from different perspectives; by connecting the resources (e.g. funding, capital resources, etc.), we can analyze the alignment between investment and revenue, and analyze the resources flow from one component to the other thereby optimize investment strategies; and by connecting data, we analyze how data being shared and used among components. 

\subsection{Impact Diffusion Analysis}
Modules in many conventional BAs internally are isolated with each other, externally are isolated from the up-to-date data including dynamic marketplaces, changing customer demands, etc. In this section, we discuss how components affect each other through the impact driven by internal and external data, which is essential for decision making, investment prioritization, and risk assessment and management.

\begin{figure}[!t]
\centering
\includegraphics[width=3.4in]{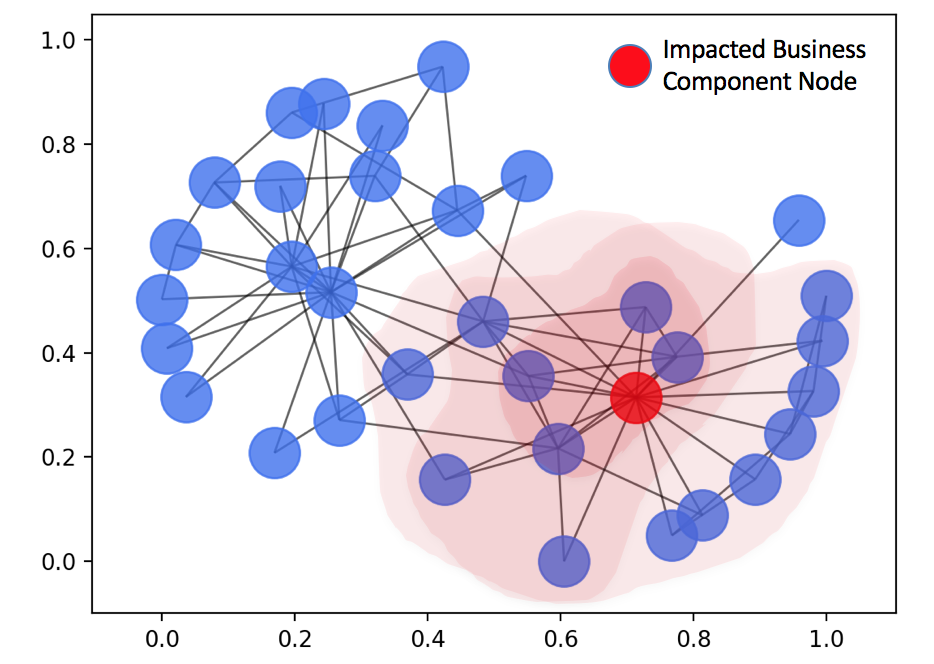}
\caption{Impact Diffusion.}
\label{fig_diffusion}
\end{figure}

In the CBM model, each component can be a target for internal or external impacts. The task of diffusion analysis is to evaluate the way of impacts diffusing from the target component to others. If we consider the target component as the start node in the CBM graph, the task can be intuitively solved by graph diffusion kernels \cite{FOUSS201253}, such as random walk with restart kernel, exponential diffusion kernel, the Laplacian exponential diffusion kernel and so on. The Fig.\ref{fig_diffusion} illustrate the idea of impact diffusion analysis.
\subsubsection{Internal Impact Diffusion}
Internally, the impacts can be generated from many aspects for a given component, such as personnel adjustment, investment, revenue gain/loss, business transformation and so on. So it is of great importance for the diffusion analysis to evaluate the scope and intensity of impacts. Based on the graph kernel, we can not only measure the probability but also the intensity of impacts starting from one component to others. For example, based on regularized Laplacian (RL) kernel defined by Eq.\eqref{RL} , we can get the inference matrix $ P $, in which $ P_{i,j} $ indicates the probability of impacts starting from component $ i $ to $ j $. For Eq.\eqref{RL}, $ L = D - A $ is the Laplacian matrix made of the graph adjacent matrix $ A $ and the degree matrix $ D $, and $ \rho(L) $ is the spectral radius of $ L $ with $ 0 < \alpha < \rho(L)^{-1} $.
\begin{equation}\label{RL}%
\small
\textit{RL} =  \sum\limits_{k=0}^\infty \alpha^k{(-L)}^k = {(I+\alpha*L)}^{-1}
\end{equation} 
\subsubsection{External Impact Diffusion}
As we know that conventional BAs cannot capture external impacts well. To connect the graphical CBM model with the external data, we build a RSS news feeds for each component. Firstly, for each component, we generate feature tags based on its text information that includes title, description, processes, etc.; Then, we use those feature tags as a query to get up-to-date data from the RSS news feed; After that, we process and analyze the external data based on text mining and machine learning techniques to obtain potential external impacts; and those impacts can be further ranked and labeled based on their importance and sentiment analysis, namely positive, neutral and negative. Similarly, we can adopt various graph diffusion kernels to measure how far and how much the impact from the start node will affect other nodes. The external RSS news feeds enable us to build a more automatic pipeline. But the external impacts can be generated from any external sources that are not in BAs, which definitely are worthy to be investigated.

\section{Conclusion}
In this work we analyzed some major limitations of conventional BAs, including inflexible structure, isolated BA modules, and lack of data-driven analytics. To solve those problems, we proposed the graph theory based data-driven analytics with aim of automatically and intelligently generating up-to-date business insights. We illustrated the proposed ideas with the conventional CBM model as our case study and implemented a CBM graph for the data-driven analytics. We demonstrated the CBM graph can effectively serve as bridges to internally connect CBM components, and externally connect to outside data and impacts. We analyzed potential applications of our method from different perspectives, which include structure visualization, business insights based on graph topological features, and internal and external impact diffusion.

In the future, we will start to implement the proposed ideas with actual CBM models and enterprise data. Building a robust CBM platform with graph theory based data-driven analytics definitely needs systematic study and development. Specifically, main tasks may include: a) redesign the back-end for CBM graph to make the platform scalable, and get the graph theory based analytics done efficiently. Given the traditional database may not satisfy those requirements well, graph database may be needed for the new BA; b) redesign the front-end to show various analytics results more user-friendly and effectively; c) we also need to evaluate and refine internal and external data sources so as to generate more accurate business insights. We believe our graph theory based data-driven analytics can be a promising part for enhancing business intelligence. So, in the meantime, we will continue to explore additional ways in which graph theory can be leveraged for CBM and other BA models.

\bibliographystyle{IEEEtran}
\bibliography{IEEEexample.bib}

\end{document}